\documentclass[prd,twocolumn,showpacs,preprintnumbers,amsmath,amssymb,superscriptaddress,floatfix,nofootinbib]{revtex4}

\usepackage{graphicx}
\usepackage{epstopdf}
\usepackage{bm}
\usepackage{amsmath}
\usepackage{amsfonts}
\usepackage{amssymb}
\usepackage{color}
\usepackage{multirow}
\usepackage{footnote}

\usepackage[colorlinks, citecolor=blue,anchorcolor=red,menucolor=red,linkcolor=red,filecolor=red,runcolor=red,urlcolor=blue,frenchlinks=red]{hyperref}
\begin{document}
\title{Spectrum and decay properties of the charmed mesons involving the coupled channel effects}
\author{Wei Hao}
\email{haowei@nankai.edu.cn}
\affiliation{School of Physics, Nankai University, Tianjin 300071, China}

\author{M. Atif Sultan}%
\email{atifsultan.chep@pu.edu.pk}
\affiliation{School of Physics, Nankai University, Tianjin 300071, China}
\affiliation{Centre  For  High  Energy  Physics,  University  of  the  Punjab,  Lahore  (54590),  Pakistan}

\author{En Wang}~\email{wangen@zzu.edu.cn}
\affiliation{School of Physics, Zhengzhou University, Zhengzhou 450001, China}

\begin{abstract}
The mass spectrum of the charmed mesons is investigated by considering the coupled channel effects within the nonrelativistic potential model. The predicted masses of the charmed mesons are in agreement with experimental data. The strong decay properties are further analyzed within the $^3P_0$ model by using numerical wave functions obtained from nonrelativistic potential model. Based on the predicted masses and decay properties, we give a classification of the recently observed charmed states. Especially, we have effectively explained the masses and decay properties of the $D_1^*(2600)$ and $D_1^*(2760)$ by considering the $S$-$D$ mixing. Furthermore, the predicted masses and decay properties of the $2P$ wave states are helpful to search for them experimentally in future.

\end{abstract}

\maketitle

\section{Introduction}
In the traditional quark model~\cite{Gell-Mann:1964ewy,Zweig:1964ruk},  the hadrons are composed of constituted quarks, i.e. the $q\bar{q}$ mesons and the $qqq$ baryons. While most of the observed hadrons can be well described within the traditional quark models, there are still many hadrons with exotic properties that can not be explained within the traditional quark models, which implies that the hadrons have more complex structure, and it is necessary to further reveal their internal information by considering the unquenched effect~\cite{Wang:2024jyk}. Within the Quantum chromodynamics (QCD), the high Fock components in hadron wave functions should be existed~\cite{Bali:2005fu,Armoni:2008jy,Bigazzi:2008gd,PACS-CS:2011ngu}, and the unquenching lattice QCD effect can be achieved through hadron self energy, which shifts the hadron masses~\cite{Heikkila:1983wd,Barnes:2007xu,Pennington:2007xr}.

\begin{table}[!htpb]
\begin{center}
\caption{ \label{exp.cu} Masses, widths, quantum numbers, and the strong decay modes of the charmed mesons~\cite{ParticleDataGroup:2024cfk}.}
\footnotesize
\resizebox{\linewidth}{!}{
\begin{tabular}{lcccc}
\hline\hline
  state                 &mass  (MeV)                     &width  (MeV)           & $J^{P}$    &decay modes        \\\hline
  $D^\pm$                   &$1869.66\pm0.05$                &$-$                    & $0^-$    &$-$       \\
  $D^0$                     &$1864.84\pm0.05$                &$-$                    & $0^-$    &$-$       \\
  $D^*(2007)^0$             &$2006.85\pm0.05$                &$<2.1$                 & $1^-$    &$D^0\pi^0,D^0\gamma$ \\
  $D^*(2010)^\pm$           &$2010.26\pm0.05$                &$0.0834\pm0.0018$      & $1^-$    &$D^0\pi^+,D^+\pi^0$ \\
  $D_0^*(2300)$             &$2343\pm10$                     &$229\pm16$             & $0^+$    &$D\pi^\pm$       \\
  $D_1(2420)$               &$2422.1\pm 0.6$                 &$31.3\pm1.9$           & $1^+$    &$D^{*0}\pi$     \\
  $D_1(2430)^0$             &$2412\pm 9$                     &$314\pm29$             & $1^+$    &$D^{*+}\pi^-$\\
  $D^*_2(2460)$             &$2461.1^{+0.7}_{-0.8}$          &$47.3\pm0.8$           & $2^+$    &$D\pi^-,D^*\pi^-$     \\
  $D_0(2550)^0$             &$2549\pm19$                     &$165\pm24$             & $0^-$    &$D^{*+}\pi^-$\\
  $D_1^*(2600)^0$           &$2627\pm10$                     &$141\pm23$             & $1^-$    &$D\pi,D^*\pi$  \\
  $D^*(2640)^\pm$        &$2637\pm6$                      &$<15$                  & $?^?$    &$D^{*+}\pi^+\pi^-$  \\
  $D_2(2740)^0$             &$2747\pm6$                      &$88\pm19$              & $2^-$    &$D^{*+}\pi^-$ \\
  
  $D_3^*(2750)$             &$2763.1\pm 3.2$                 &$66\pm5$               & $3^-$    &$D\pi,D^*\pi$    \\
  $D_1^*(2760)^0$           &$2781\pm22$                     &$180\pm40$             & $1^-$    &$D^+\pi^-$ \\
  $D(3000)^0$               &$3210\pm60$                     &$190\pm80$             & $?^?$    &$D^{*+}\pi^-$\\
  \hline\hline
\end{tabular}}
\end{center}
\end{table}

The charmed meson family was first experimentally observed in 1976~\cite{Goldhaber:1976xn,Peruzzi:1976sv,Wiss:1976gd}. Up to now, the properties of ground charmed mesons have been well understood. The BaBar, Belle, CLEO, BESIII, and LHCb experiments have discovered numerous excited charmed states. The $1S$ states ($D$ and $D^*$) and $1P$ states ($D_0^*(2400)$, $D_1(2420)$, $D_1(2430)$, and $D_2^*(2460)$) have been well established. However, there are several charmed mesons, i.e. $D_0(2500)^0$, $D_1^*(2600)^0$, $D^*(2640)^\pm$, $D_2(2740)^0$, $D_3^*(2750)$, $D_1^*(2760)^0$, and $D(3000)^0$, which have not  been well described~\cite{ParticleDataGroup:2024cfk}, and we have listed the experimental information about those charmed mesons in Table~\ref{exp.cu}.

The mass spectrum of the charmed mesons has been investigated by many theoretical groups with various models. 
As the first principles, lattice QCD has conducted a series of studies on the spectrum of the charmed mesons~\cite{Moir:2013ub,Lewis:2000sv,Cichy:2015tma}. In addition, there are studies of the charmed meson spectrum within the phenomenological models, such as the Godfrey-Isgur (GI) model~\cite{Godfrey:1985xj,Godfrey:2015dva}, the screened GI model~\cite{Song:2015fha}, the QCD-motivated relativistic quark model based on the quasi-potential approach~\cite{Ebert:2009ua}, the heavy meson effective theory~\cite{Wang:2013tka}, and the nonrelativistic constituent quark model~\cite{Li:2010vx}.

Despite numerous attempts to explain and predict the properties of charmed mesons, there are still significant discrepancies between theoretical predictions and experimental results regarding the masses of multiple charmed mesons. For instance, Ref.~\cite{Song:2015fha} has provided a good explanation for the suppression effect of the excited charmed meson masses by introducing the screened effects. It is shown that the coupled channel effects play an advantage in the interpretation of hadron states, such as the $X(3872)$~\cite{Kalashnikova:2005ui} and $D_{s0}(2317)$~\cite{Yang:2021tvc}.
Therefore, investigating the spectrum of the charmed mesons involving the coupled channel effects could improve our understanding the properties of the observed charmed mesons, and provide more information for experimental searches for them. 

Coupled channel effects have been studies many years ago~\cite{Heikkila:1983wd, Ono:1983rd,Tornqvist:1984fy,Ono:1985eu,Ono:1985jt}, and currently it is still an important tool for investigating hadron spectrum~\cite{Kalashnikova:2005ui,Li:2009ad,Ferretti:2012zz,Ferretti:2013vua,Ferretti:2013faa,Ortega:2016mms,Lu:2016mbb,Ortega:2021fem,Hao:2022vwt,Yang:2023tvc}.
The coupled channel model is an unquenched quark model which  considers the virtual hadronic loop effects in the quenched quark model.
The basic assumption of this model is the mixing between the $q\bar{q}$ component and the two-meson component in a hadron state. The hadronic loop effects have been proven to be highly nontrivial, and it can contribute continuum components to the physical hadron state and lead to mass shift of the bare component ($q\bar{q}$)~\cite{Liu:2011yp}. 
This effect can also be called pair-creation effect, which describes the coupling between bare state and meson-meson (meson-baryon) channels.
The coupling can lead to strong decay when the bare mass is larger than intermediate state threshold.
By considering the coupled channel effect, we have studied the mass spectrum and decay properties of the charmed-strange mesons~\cite{Hao:2022vwt,Yang:2023tvc}.

In this work, we would like to investigate the mass spectrum of the charmed mesons by considering the coupled channel effects and the strong decays within the nonrelativistic quark model~\cite{Li:2010vx,Lu:2016bbk}. 
This paper is organized as follows. In Section~\ref{sec:formalism}, we present the theoretical framework of the nonrelativistic quark model, and the coupled channel model. In Section~\ref{sec:results}, we show our calculated results of the charmed meson masses including coupled channel effects and strong decays, and compare our results with the experimental data. Finally, in Section~\ref{sec:summary}, we present a short summary of this work.

\section{Theoretical Formalism}\label{sec:formalism}

\subsection{Nonrelativistic quark model}
The nonrelativistic quark model is one of the quenched quark models, which represents the quenched effects of a coupled channel system. This model has been widely used in the study of hadron properties such as bottom mesons~\cite{Lu:2016bbk,Feng:2022esz,Hao:2022ibj}, open charmed mesons~\cite{Li:2010vx}, and charmonium~\cite{Barnes:2005pb,Hao:2019fjg}. The Hamiltonian $H_A$ of the model includes the Coulomb term, liner term, and spin-dependent term that reflects the interactions of the quark and antiquark of a meson~\cite{Lakhina:2006fy}, which can be expressed as
\begin{align} \label{eq:hamilt}
    H_A =& H_0+H_{sd},\\
    H_0 =& m_{q} + m_{\bar{q}} + \frac{\boldsymbol{P}^2}{2M_r}-\frac{4}{3}\frac{{\alpha}_s}{r}+br+C_{q\bar{q}}\nonumber\\  & + \frac{32{\alpha}_s{\sigma}^3 e^{-{\sigma}^2r^2}}{9\sqrt{\pi}m_qm_{\bar{q}}} {\boldsymbol{S}}_{q} \cdot {\boldsymbol{S}}_{\bar{q}},
\end{align}
where $m_{q}$, $m_{\bar{q}}$, $\alpha_s$, $b$, $C_{q\bar{q}}$, and $\sigma$ are the free parameters that need to refit the masses of the well-established charmed states. $M_r=m_q m_{\bar{q}}/(m_q+m_{\bar{q}})$ is the reduced mass. $\boldsymbol{S_q}$ and $\boldsymbol{S_{\bar{q}}}$ are the spin of the quark and antiquark of the meson. 

The spin-dependent term $H_{sd}$ can be written as
    \begin{eqnarray}
      H_{sd} &=& \left(\frac{\boldsymbol{S}_{q}}{2m_q^2}+\frac{{\boldsymbol{S}}_{\bar{q}}}{2m_{\bar{q}}^2}\right) \cdot \boldsymbol{L}\left(\frac{1}{r}\frac{dV_c}{dr}+\frac{2}{r}\frac{dV_1}{dr}\right)\nonumber\\
      &&+\frac{{\boldsymbol{S}}_+ \cdot \boldsymbol{L}}{m_qm_{\bar{q}}}\left(\frac{1}{r} \frac{dV_2}{r}\right) \nonumber\\
      && +\frac{3{\boldsymbol{S}}_{q} \cdot \hat{\boldsymbol{r}}{\boldsymbol{S}}_{\bar{q}} \cdot \hat{\boldsymbol{r}}-{\boldsymbol{S}}_{q} \cdot {\boldsymbol{S}}_{\bar{q}}}{3m_qm_{\bar{q}}}V_3\nonumber\\
      && +\left[\left(\frac{{\boldsymbol{S}}_{q}}{m_q^2}-\frac{{\boldsymbol{S}}_{\bar{q}}}{m_{\bar{q}}^2}\right)+\frac{{\boldsymbol{S}}_-}{m_qm_{\bar{q}}}\right] \cdot \boldsymbol{L} V_4,
\end{eqnarray}
where the $r=|\bold{r}|=|\bold{r_q}-\bold{r_{\bar{q}}}|$ is the separation of two quarks in a meson. $\boldsymbol{L}$ is the orbital angular momentum between quark $q$ and antiquark $\bar{q}$. $\boldsymbol{S}_{\pm}={\boldsymbol{S}}_q\pm{\boldsymbol{S}}_{\bar{q}}$.  $V_i$ are the Wilson loop matrix elements, which are given as
\begin{eqnarray}
  V_c &=& -\frac{4}{3}\frac{{\alpha}_s}{r}+br,\nonumber \\
  V_1 &=& -br-\frac{2}{9\pi}\frac{{\alpha}_s^2}{r}\left[9{\rm ln}\left(\sqrt{m_qm_{\bar{q}}}r\right)+9{\gamma}_E-4\right],\nonumber\\
  V_2 &=& -\frac{4}{3}\frac{{\alpha}_s}{r}-\frac{1}{9\pi}\frac{{\alpha}_s^2}{r}\left[-18{\rm ln}\left(\sqrt{m_qm_{\bar{q}}}r\right)+54{\rm ln}(\mu r)\right.\nonumber\\
  &&\left.+36{\gamma}_E+29\right],\nonumber\\
  V_3\ &=& -\frac{4{\alpha}_s}{r^3}-\frac{1}{3\pi}\frac{{\alpha}_s^2}{r^3}\left[-36{\rm ln}\left(\sqrt{m_qm_{\bar{q}}}r\right)+54{\rm ln}(\mu r) \right.\nonumber\\
  &&\left.+18{\gamma}_E+31\right],\nonumber\\
  V_4 &=& \frac{1}{\pi}\frac{{\alpha}_s^2}{r^3}{\rm ln}\left(\frac{m_{\bar{q}}}{m_q}\right),
\end{eqnarray}
where $\gamma_E$ is Euler constant, and $\mu$ is renormalization scale. 
The value of the parameters are taken from Ref.~\cite{Lakhina:2006fy}, i.e. $\gamma_E=0.5772$ and $\mu=1$~GeV.

With the Hamiltonian of Eq.~(\ref{eq:hamilt}), the mixing angle can be estimated by analysing the spin-orbit term $H_{sd}$. 
The spin-orbit term can be divided into symmetric part $H_{sym}$ and antisymmetric part $H_{anti}$~\cite{Lu:2016bbk},
\begin{eqnarray}
H_{sym} &=& \frac{{\boldsymbol{S}}_+ \cdot {\boldsymbol{L}}}{2}\left[\left(\frac{1}{2m_q^2}+\frac{1}{2m_{\bar{q}}^2}\right) \left(\frac{1}{r}\frac{dV_c}{dr}+\frac{2}{r}\frac{dV_1}{dr}\right)\right. \nonumber \\
&& \left.+\frac{2}{m_qm_{\bar{q}}}\left(\frac{1}{r} \frac{dV_2}{r}\right)+\left(\frac{1}{m_q^2}-\frac{1}{m_{\bar{q}}^2}\right)V_4\right],
\end{eqnarray}
\begin{eqnarray}
\label{mix}
H_{anti} &=& \frac{{\boldsymbol{S}}_- \cdot {\boldsymbol{L}}}{2}\left[\left(\frac{1}{2m_q^2}-\frac{1}{2m_{\bar{q}}^2}\right) \left(\frac{1}{r}\frac{dV_c}{dr}+\frac{2}{r}\frac{dV_1}{dr}\right)\right. \nonumber \\
&& \left.+\left(\frac{1}{m_q^2}+\frac{1}{m_{\bar{q}}^2}+\frac{2}{m_qm_{\bar{q}}}\right)V_4\right].
\end{eqnarray}
The antisymmetric part $H_{anti}$ gives rise to the the spin-orbit mixing of the mesons by providing the influence of non-diagonal terms, and the mixing angle can be extracted through the diagonalization. 
This mixing can be parametrized by a mixing angle $\theta_{nL}$ as follows~\cite{Godfrey:1985xj,Godfrey:1986wj,Lu:2016bbk},
\begin{equation}
\left(
\begin{array}{cr}
D_{L}(nL)\\
D^\prime_{L}(nL)
\end{array}
\right)
 =\left(
 \begin{array}{cr}
\cos \theta_{nL} & \sin \theta_{nL} \\
-\sin \theta_{nL} & \cos \theta_{nL}
\end{array}
\right)
\left(\begin{array}{cr}
D(n^1L_L)\\
D(n^3L_L)
\end{array}
\right),
\label{eqn:mix}
\end{equation}
where $D_{L}(nL)$ and $D_{L}^\prime(nL)$ represent the physical observed states.

\subsection{Coupled Channel Model}
In the coupled channel framework, the full Hamiltonian consists of two parts, one is the quenched part and another is the coupled channel part which is,
\begin{equation} \label{eqn:full}
	H = H_A + H_{BC} + H_I,
\end{equation}
where $H_A$ is the quenched part achieved by the non-relativistic quark model, as shown in Eq.~(\ref{eq:hamilt}), $H_{BC}$ is the Hamiltonian between the meson pairs, and $H_I$ denote the mixing between $c\bar{u}$ bare state and $BC$ meson pair system which is accomodated by the $^3P_0$ model. The $H_{BC}$ is expressed as,
\begin{align}
    H_{BC} = E_{BC} =\sqrt{m_B^2 +p^2} + \sqrt{m_C^2 +p^2}.
\end{align}

For the term $H_I$, we use the widely used $^3P_0$ model to reflect the coupled channel effects \cite{Micu:1968mk, LeYaouanc:1972vsx, LeYaouanc:1973ldf}.
This model is based on the basic assumption that $q\bar{q}$ pairs can combine with vacuum generated $q\bar{q}$ pairs to form two meson pairs.
The generated quark-antiquark pairs have vacuum quantum numbers $J^{PC} = 0^{++}$ which mean the spin and orbital angular momentum of the quark-antiquark pair is 1 and the spectroscopy notation $^{2S +1}L_J$ of the system reads $^3P_0$.

In the $^3P_0$ model, the  operator $T^\dag$ is of quark-antiquark pair-creation, as given by~\cite{Ferretti:2013faa,Ferretti:2012zz,Ferretti:2013vua}
\begin{equation}
	\label{eqn:Tdag}
	\begin{array}{rcl}
	T^{\dagger} &=& -3 \, \gamma_0^{eff} \, \int d \vec{p}_3 \, d \vec{p}_4 \, 
	\delta(\vec{p}_3 + \vec{p}_4) \, C_{34} \, F_{34} \,  
	{e}^{-r_q^2 (\vec{p}_3 - \vec{p}_4)^2/6 }\,  \\
	& & \left[ \chi_{34} \, \times \, {\cal Y}_{1}(\vec{p}_3 - \vec{p}_4) \right]^{(0)}_0 \, 
	b_3^{\dagger}(\vec{p}_3) \, d_4^{\dagger}(\vec{p}_4) .
	\end{array}
\end{equation}
It should be pointed that the operator $T^{\dagger}$ includes a Gaussian factor ${e}^{-r_q^2 (\vec{p}_3 - \vec{p}_4)^2/6 }$, which is adopted to smear out the pair-creation point~\cite{Ferretti:2013faa,Ferretti:2012zz,Ferretti:2013vua,Silvestre-Brac:1991qqx,Geiger:1991ab,Geiger:1991qe,Geiger:1996re}, where the parameter $r_q$ represents the effective size of the created quark pair. 
The value of $r_q$ was determined from meson decays to be in the range $0.25\sim 0.35$~fm~\cite{Silvestre-Brac:1991qqx,Geiger:1991ab,Geiger:1991qe,Geiger:1996re}. In our calculation, we take the value $r_q = 0.3$~fm, as used in Refs.~\cite{Yang:2023tvc,Hao:2022vwt}. 
In Eq.~(\ref{eqn:Tdag}), the color, flavor and spin wave functions of the created quark-antiquark system are labeled as $C_{34}$, $F_{34}$, and $\chi_{34}$, respectively.  
$ b_3^{\dagger}(\vec{p}_3)$ and $d_4^{\dagger}(\vec{p}_4)$ are the creation operators for a quark and an antiquark with momenta $\vec{p}_3$ and $\vec{p}_4$, respectively.  
$\gamma_0^{eff}=\frac{m_n}{m_i}\gamma_0$ is the pair-creation strength, where $m_n$ represents the light quark mass $m_{u(d)}$, and $m_i$ represents to the quark mass ($i=u,d,s$).  As we know $\gamma_0^{eff}=0.4$ is a  typical value to calculate strong decay properties of a meson system. However, for different meson system, its value usually different~\cite{Segovia:2012cd}.
Thus in this work, we obtain the $\gamma_0$ by fitting the decay width of the $D_2^*(2460)$ which can be well assigned as $D(1^3P_2)$~\cite{Li:2010vx,Pandya:2021ddc,Gupta:2018zlg}. 

By considering the quenched part and the coupled channel part of the Hamiltonian of Eq.~(\ref{eqn:full}), the full eigen function can be represented by
\begin{align} \label{eqn:psi}
    |\psi\rangle = c_0 |\psi_0\rangle + \sum_{BC} \int d^3p\, c_{BC}(p) |BC;p\rangle,
\end{align}
where $c_0$ and $c_{BC}(p)$ are the normalization constants of the two components for a meson system, and $p$ is the momentum of meson $B$ in the rest frame of meson $A$ . In this scheme, the mass of physical mass $M$ of a conventional meson is given by
\begin{align}
\label{m}
M &= M_0 + \Delta M, \\
\Delta M &= \sum_{BC} \int_0^{\infty} d^3p \frac{\left|\left\langle BC;p \right| T^\dagger \left| \psi_0 \right\rangle \right|^2}{M - E_{BC}},
\end{align}
where the sum runs over all meson pairs that couple to meson $A$. The physical mass include two parts, the first one $M_0$ is the eigenvalue of the quenched Hamiltonian $H_A$ and $\Delta M$ represents the mass shift caused by the coupled channel effects.

As we know, the coefficient $c_0$ and $c_{BC}(p)$ in Eq.~(\ref{eqn:psi}) is related with  probabilities of the components in the meson system. For the state below the $BC$ threshold, it is valid to normalize the physical state $|\psi\rangle$, and the probability of quenched quark pairs can be calculated as
\begin{equation}
\label{eqn:pqqbar}
	\mathcal{P}_{Q\bar{q}} \equiv |c_0|^2 = \left(1+\sum_{BC} \int_0^{\infty} d^3p \frac{\left|\left\langle BC;p \ell J \right| T^\dagger \left| \psi_0 \right\rangle \right|^2}{(M - E_{BC})^2}\right)^{-1},
\end{equation}
where the $BC$ component can be written  as $P_{BC}= 1- P_{Q\bar{q}}$.

Besides, for state $A$ above the threshold of $B$ and $C$, the strong decay will occur, and the width can also be calculated with the following formula,
\begin{equation}
\label{eqn:decay}
    \Gamma_{BC} = 2 \pi p_0 \frac{E_B(p_0) E_C(p_0)}{M_A} \left| \left\langle BC;p_0\right| T^\dagger \left| \psi_0 \right\rangle \right|^2,
\end{equation}
 where $p_0$ and $E_{B,C}$ are the momentum and energy of meson $B$ or $C$ in the rest frame of initial meson $A$,
\begin{gather}
p_0=\frac{\sqrt{\left[m_A^2-(m_B+m_C)^2\right]\left[m_A^2-(m_B-m_C)^2\right]} }{2m_A},\\
 E_B(p_0) = \sqrt{m_B^2 + p_0^2},\\   
E_C(p_0) = \sqrt{m_C^2 + p_0^2}.
\end{gather}

\section{Results and discussions}
\label{sec:results}
In this section, we will calculate the masses and the strong decay widths of the charmed mesons with above ingredients. 
As we know, among the charmed mesons, only a few states are well established as pure $q\bar{q}$ components. Thus, we obtained the values of the parameters by fitting the masses of the well assignment states $D(1^1S_0)$, $D^*(1^3S_1)$, $D_2^*(2460)(1^3P_2)$, and $D_3^*(2750)(1^3D_3)$, and the decay width of $D_2^*(2460)(1^3P_2)$, and tabulated those parameters in Table~\ref{tab:para}. 
It should be noted that, in our calculation of the $^3P_0$ model, we use the realistic wave functions by solving the nonrelativistic quark model, rather than the harmonic oscillator wave function. 

Using the parameters listed in Table~\ref{tab:para},  
we have calculated the masses of the charmed mesons in coupled channel model, and the  corresponding mass shifts, as shown in Table~\ref{tab:spectrum} and Table~\ref{tab:shift}, respectively. For comparison, the  predicted masses of other  (nonrelativistic and relativistic) quark models and also the experimental data are listed in Table~\ref{tab:spectrum} as well. It is found that those well-established states can also be reasonably explained within the coupled channel model. According to Table~\ref{tab:spectrum},  our results are in good agreement with the experimental data. 

Then we have calculated the decay widths of the charmed mesons, as show in Table~\ref{tab:width1} and ~\ref{tab:width2}, respectively.

As we know, when the heavy-light mesons have different total spins but with same total angular momentum, the spin-orbit mixing will happen, and the mixing angle is important to explain the meson properties. With Eq.~(\ref{mix}),  the mixing angles of $1P$, $2P$, and $1D$ can be extracted to be $-30.0^\circ$, $-33.2^\circ$, and $-41.1^\circ$, respectively, which  are close to $\theta(1P)=-34.0^\circ$, $\theta(2P)=-23.5^\circ$, and $\theta(1D)=-40.3^\circ$ of Ref.~\cite{Ni:2021pce}, and $\theta(1P)=-25.68^\circ$, $\theta(2P)=-29.39^\circ$, and $\theta(1D)=-38.17^\circ$ of Ref.~\cite{Godfrey:2015dva}.

\subsection{$S$-wave states}
For the charmed mesons, the $S$-wave doublet $D(0^-) $, $D^*(1^-)$  have been well-established. Our results can  well reproduce the masses of the two $S$-wave states. Because the mass of $D$ is below all the channel thresholds, we can estimate probabilities of its each component. Our result show that, the charmed meson $D(1^1S_0)$  is constituted with  $79.2\%$ $c\bar{u}$ and $20.8\%$ coupled channel components as shown in Table~\ref{tab:pro}. The various coupled channel components of the $D$ meson are $D^*\pi (5.3\%)$, $D^*\eta (0.4\%)$, $D^*\eta^\prime (0.6\%)$, $D\rho (3.2\%)$, $D^*\rho (7.7\%)$, $D\omega (1.1\%)$ and $D^*\omega (2.5\%)$. 
\par
For the $2S$-wave mesons, there are two candidates $D_0(2500)^0$ and $D_1^*(2600)^0$ can be assigned. They are generally assumed to be the $D(2^1S_0)$ and $D(2^3S_1)$~\cite{Li:2022vby,Chen:2011rr,Lu:2014zua}. The $D_0(2500)$ was observed by the BaBar Collaboration in $D^*\pi$ mass distribution in the inclusive process $e^+e^-\to c\bar{c}$ in 2010~\cite{BaBar:2010zpy}, and its mass and width are measured to be $M=2539.4\pm4.5\pm6.8$~MeV and $\Gamma=130\pm12\pm13$~MeV. The data are updated by the LHCb Collaboration in 2013 through the process $pp\to D^{*+}\pi^- X$ in the $D^{*+}\pi^-$ invariant mass spectrum with $M=2579.5\pm3.4\pm5.5$~MeV and $\Gamma=177.5\pm17.8\pm46.0$~MeV~\cite{LHCb:2013jjb}. Furthermore, the LHCb updated again the experimental information in 2020 and gave its mass $M=2513\pm 2\pm7$~MeV and width $199\pm5\pm17$~MeV~\cite{LHCb:2019juy}. As we see, 
there are some discrepancies between these experimental data, and our predicted mass and width are in fair agreement with the LHCb results.

Meanwhile, the $D_1^*(2600)^0$ was first reported by the BaBar Collaboration in $D^{(*)}\pi$ mass distribution, and its mass and width are $M=2608.7\pm2.4\pm2.5$~MeV and $\Gamma=93\pm6\pm13$~MeV~\cite{BaBar:2010zpy}. Latter another state was observed by LHCb in 2013 with mass  $M=2649.2\pm2.5\pm3.5$~MeV and width $\Gamma=140.2\pm17.1\pm18.6$~MeV~\cite{LHCb:2013jjb}, which  can be regarded as the  $D_1^*(2600)^0$. 
In addition, in 2016 the LHC Collaboration observed one state with  mass 
$2681.1\pm5.6\pm14.0$~MeV and width $186.7\pm8.5\pm11.9$~MeV~\cite{LHCb:2016lxy}, which are larger than the previous results. The discrepancies between these measurements confuses our understanding of this states, and it implies that the $S$-$D$ mixing of the $D(2^3S_1)$ and $D(1^3D_1)$ should be considered, which we will discuss later.

\subsection{$P$-wave states}
For the $P$-wave charmed mesons, there are four states, i.e. $D_0^*(2300)$, $D_1(2420)$, $D_1(2430)$, and $D_2^*(2460)$. The mass range of the $D_0^*(2300)$ is about $2.29\sim 2.4$ GeV~\cite{ParticleDataGroup:2024cfk}. Our predicted mass (2261~MeV) is a little smaller than the experimental data but consistent with the theoretical predictions of the nonrelativistic quark model~\cite{Li:2010vx} and the relativistic quark model~\cite{Zeng:1994vj}. We have predicted the dominant decay mode $D\pi$,  same as Ref.~\cite{Song:2015fha}, but the predicted total width (159~MeV) is smaller than experimental result $229\pm16$~MeV~\cite{ParticleDataGroup:2024cfk}. It should be stressed that, there are some puzzle about the $D_0^*(2300)$ and its strange partner state $D_{s0}^*(2317)$~\cite{Albaladejo:2016lbb,Du:2019oki,Lyu:2023ppb,Liu:2022dmm}. If we regard $D_{s0}^*(2317)$ as $D_s(1^3P_0)$, it is impossible to assign $D_0^*(2300)$ as $D(1^3P_0)$, because the strange $(s)$ quark is heavier than the up $(u)$ quark. However, according to the existing experimental information, $D_0^*(2300)$ has a larger mass than $D_{s0}^*(2317)$. As we know, although the $D_{s0}^*(2317)$ has some exotic state explanations such as hadronic molecules and compact tetraquark states~\cite{Barnes:2003dj,Browder:2003fk,Lipkin:2003zk,Bicudo:2004dx,Dmitrasinovic:2005gc,Yang:2021tvc,Liu:2022dmm}, the traditional explanation of meson states cannot be completely ruled out~\cite{Lakhina:2006fy,vanBeveren:2003kd,Dai:2006uz,Liu:2009uz,Hao:2022vwt}. Thus there is still significant controversy between the experimental and theoretical explanations of these two states, and more investigation is needed. 

The $D_1(2420)$ and $D_1(2430)$ are the mixtures of the $1^1P_1$ and $1^3P_1$ states. We estimate the mixing angle $\theta_{1P}=-30.0^\circ$, which is close to $\theta_{1P}=-25.68^\circ$ obtained in the relativized quark model~\cite{Godfrey:2015dva}. With this angle, the predicted masses of the two states are 2378~MeV and 2454~MeV. 
For the decay widths of the two states, the width range of $D_1(2420)$ is about $13\sim 58$~MeV~\cite{ARGUS:1989mcc,TaggedPhotonSpectrometer:1988qan,ParticleDataGroup:2024cfk}, and the width range of $D_1(2430)$ is about $266\sim384$~MeV. We predict that the width of $D(1^1P_1)$ is 13~MeV, which is consistent with the lower limit of the experimental result. However the predicted width (162~MeV) of $D_1(1^3P_1)$  is smaller than experimental data. Of course, the decays of these two states is related to the mixing angle. In our calculation, we obtain the mixing angle by analyzing anti-symmetric part of the the spin-orbit term. This mixing angle is influenced by the model parameters. It should be noted that, as shown in Table~\ref{exp.cu}, there is still significant uncertainties in the experimental measurements about the $D$ mesons.

As for the $D(1^3P_2)$ state, by considering the mass and width, the $D_2^*(2460)$ can be a good candidate.  From Table~\ref{tab:width2}, we predicted the $D(1^3P_2)$ mainly decay to $D\pi$ and $D^*\pi$ with width 29~MeV and 21~MeV, respectively. Besides, we calculated the decay branching fraction  
\begin{equation}
  \frac{\Gamma(D_2^*(2460)\to D^+\pi^-)}{\Gamma(D_2^*(2460)\to D^{*+}\pi^-)}=1.39
\end{equation}
which is good agreement with the experiment data $1.4\pm0.3\pm0.3$ from ZEUS in 2013~\cite{ZEUS:2012gyr}, and $1.47\pm0.03\pm0.16$ from BaBar in 2010~\cite{BaBar:2010zpy}. 

\subsection{$2P$-wave states}
For the $2P$-wave mesons, there is no experimental information up to now.  We predicted the masses of the $D(2^3P_0)$, $D(2P)$, $D(2P^\prime)$, and $D(2^3P_2)$, which are 2816~MeV, 3131~MeV, 3234~MeV, and 3230~MeV, respectively.   The $D(2^3P_0)$ mainly decays to $D\pi$, $D^*\rho$, $D^*\omega$, $D_1(2420)\pi$, and $D_sK$, and the total width is $213$~MeV. The $D(2P)$ and $D(2P^\prime)$ are the mixing states of $D(2^3P_1)$ and $D(2^1P_1)$, and the mixing angle is obtained to be $-33.2^\circ$. With the mixing angle, we predicted that the $D(2P)$ dominantly decays to $D^*\pi$ $D\rho$, $D^*\rho$, $D^*\omega$, $D_2^*(2460)\pi$, $D_sK^*$, $D_s^*K^*$, and $D(2P^\prime)$ mainly decays to $D\rho$, $D^*\rho$ and $D_s^*K^*$. The total widths of the two states are 428~MeV and 211~MeV, respectively. As for the $D(2^3P_2)$ state, the $D\pi$, $D^*\pi$, $D^*\rho$, $D^*\omega$, $D_2^*(2460)\pi$, $D_1(2420)\pi$, and $D_s^*K^*$ are its mainly decay modes, and the total width is about 399~MeV. By considering the various decay modes, the four $2P$-wave states can be well distinguished. Our results are comparable with various references such as Refs.~\cite{Song:2015fha,Ebert:2009ua,Wang:2013tka,Li:2010vx}, which is contributed to future observations and confirmation of these states.

\subsection{$D$-wave states}
\begin{figure}[!htpb]
  \centering
   \begin{tabular}{c}
  \includegraphics[scale=0.7]{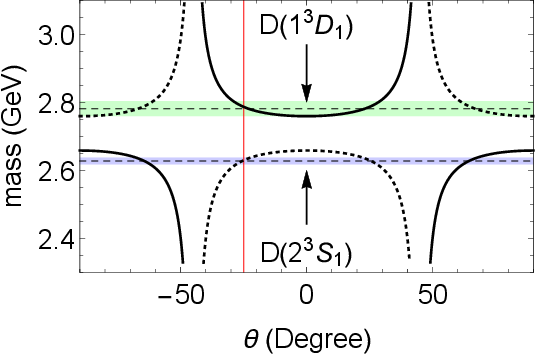}
   \end{tabular}
  \caption{Masses of the $D(1^3D_1)$ and $D(2^3S_1)$ by considering the $S$-$D$ mixing. The blue and red bands are experimental values of the two states. The red line represents the mixing angle position of $-25^\circ$. }
  \label{figms}
\end{figure}

\begin{figure}[!htpb]
  \centering
   \begin{tabular}{c}
  \includegraphics[scale=0.7]{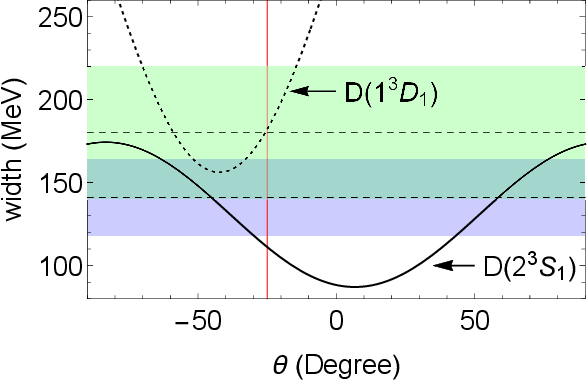}
   \end{tabular}
  \caption{Total decay widths  of the $D(1^3D_1)$ and $D(2^3S_1)$ by considering the $S$-$D$ mixing. The blue and red bands are experimental values of the two states. The red line represents the mixing angle position of $-25^\circ$.}
  \label{figdy}
\end{figure}

For the $D$-wave states, there are three candidates $D_1^*(2760)$, $D_2(2740)$,  and $D_3^*(2750)$.  The $D_1^*(2760)$ is observed by LHCb in the process $B^-\to D^+K^-\pi^-$~\cite{LHCb:2015eqv}, and its mass is $M=2781\pm22$~MeV, which is consistent with our calculation $2759$~MeV, while its experimental width $180\pm40$~MeV is consistent with our prediction $251$~MeV within $2\sigma$ uncertainties. 
As we know, the $D(2^3S_1)$ and $D(1^3D_1)$ with same $J^P=1^-$ can lead to mixing. We have plotted the dependence of the masses of $D(2^1S_0)$ and $D(1^3D_1)$ states on the mixing angle in Fig.~\ref{figms}. From the figure, we can see that, when mixing angle is about $-24^\circ\sim-26^\circ$, the masses of the $D_1^*(2600)^0$ and $D_1^*(2760)$ can be well reproduced. Meanwhile, with the same mixing angle, the predicted decay widths of these two states are consistent with the experimental data within the uncertainties, as shown in Fig.~\ref{figdy}.
\par
The $D_2(2740)$ was found by LHCb in process $pp\to D^{*+}\pi^-X$ in 2013 with mass $M=2737.0\pm3.5\pm11.2$~MeV and width $\Gamma=73.2\pm13.4\pm25.0$~MeV~\cite{LHCb:2013jjb}, which were updated to be $M=2751\pm3\pm7$~MeV and $\Gamma=102\pm6\pm26$~MeV in 2020 through the process $B^-\to D^{*+}\pi^-\pi^-$ by LHCb~\cite{LHCb:2019juy}. In our model, the mass is predicted to be $M=2732$~MeV, which is in good agreement with the experimental data,  but the predicted width $\Gamma=221$~MeV is  larger than the experimental value. 
However, it should be noticed that we just choose mixing angle $\theta_{1D}=-41.1^\circ$ of the $D(1^1D_2)$ and $D(1^3D_2)$ to calculate the decay width.  The mixing angle $\theta_{1D}=-41.1^\circ$, which is close to the modified Godfrey-Isgur (MGI) model's result $-73.8^\circ\sim-35.7^\circ$~\cite{Song:2015fha} and $\theta_{1D}=-38.17$ in a relativized quark model~\cite{Godfrey:2015dva}. The $D_3^*(2750)$ has been confirmed by many experiments~\cite{BaBar:2010zpy,LHCb:2013jjb,LHCb:2015klp,LHCb:2016lxy,LHCb:2019juy},  the ranges of its mass and width are given by $2760\sim2823$~MeV and $34\sim197$~MeV, respectively. Our predicted mass and width of this state are $M=2756$~MeV and $\Gamma=89$~MeV, which are in agreement with the experimental data. And the predicted mainly decay channels are $D\pi$ and $D^*\pi$, which are also in consistent with experimental measurements~\cite{ParticleDataGroup:2024cfk}.

\begin{table}[h] 
\caption{Potential model parameters.} 
\label{tab:para}
\begin{center}
\begin{tabular}{ccc} 
\hline 
\hline
parameters  &  this work            &Ref.~\cite{Li:2010vx} \\ 
\hline
$m_n$      & $0.45$ GeV      & $0.45$ GeV \\
$m_s$      & $0.55$ GeV      & $0.55$ GeV\\
$m_c$      & $1.43$ GeV      & $1.43$ GeV\\  
$\alpha_s$ & $0.6192$         & $0.5$  \\  
$b$        & $0.12495$ GeV$^2$ & $0.14$ GeV$^2$\\  
$\sigma$   & $1.2073$ GeV     & $1.17$ GeV \\
$C_{cs}$   & $0.21634$ GeV     & $-0.325$ GeV \\  
$\gamma_0$ & $0.668$          & $0.452$  \\
\hline 
\hline
\end{tabular}
\end{center}
\end{table}

\begin{table*}[htpb]
\begin{center}
\caption{\label{tab:spectrum} The mass spectrum (in MeV) of the charmed mesons.
Column 3 to 5 stand for spectrum from the potential model, the mass shift, the spectrum with coupled channel effects.
Results from Ref.~\cite{Li:2010vx,Song:2015fha,Ebert:2009ua,Godfrey:2015dva} are listed in Columns $6\sim9$ as comparison.
The last Column is the experimental values taken from Review of Particle Physics (RPP)~\cite{ParticleDataGroup:2024cfk}. The mixing angles of $1P$, $2P$, $1D$ are $-30.0^\circ$, $-33.2^\circ$ and $-41.1^\circ$ respectively. The numbers in [ ] represent the results by considering the $SD$ mixing angle of $-25^\circ$.
}
\footnotesize
\begin{tabular}{cccccccccc}
\hline\hline
  $n^{2S+1}L_J$  & state            &$M_0$  &$\Delta M$  &$M$  &NR~\cite{Li:2010vx}  &~MGI\cite{Song:2015fha} &RE~\cite{Ebert:2009ua} &GI~\cite{Godfrey:2015dva} & RPP~\cite{ParticleDataGroup:2024cfk}  \\\hline
  $1^1S_0$     & $D$                &$2225$   &$-360$     &$1865$   &$1867$  &1861  &1871  &1877  &$1864.84\pm0.05$    \\
  $1^3S_1$     & $D^{*}$            &$2441$   &$-434$     &$2007$   &$2010$  &2020  &2010  &2041  &$2006.85\pm0.05$    \\
  $2^1S_0$     & $D_0(2550)^0$      &$2947$   &$-395$     &$2551$   &$2555$  &2534  &2581  &2581  &$2549\pm19$           \\
  $2^3S_1$     & $D_1^*(2600)^0$    &$3051$   &$-393$ &$2658$[$2629$] &$2636$ &2632 &2593  &2643 &$2627\pm10$                      \\
  $1^3P_0$     & $D_{0}^*(2300)$    &$2614$   &$-353$     &$2261$   &$2252$  &2365  &2406  &2399  &$2343\pm10$      \\
  $1P$         & $D_{1}(2420)$      &$2800$   &$-421$     &$2378$   &$2402$  &2426  &2426  &2456  &$2422.1\pm0.6$       \\
  $1P^\prime$  & $D_{1}(2430)$      &$2862$   &$-408$     &$2454$   &$2417$  &2431  &2469  &2467  &$2412\pm9$     \\
  $1^3P_2$     & $D_{2}^*(2460)$    &$2925$   &$-457$     &$2468$   &$2466$  &2468  &2460  &2502  &$2461.1^{+0.7}_{-0.8}$     \\
  $2^3P_0$     & $-$                &$3103$   &$-287$     &$2816$   &$2752$  &2856  &2919  &2931  &$-$ \\
  $2P$         & $-$                &$3263$   &$-132$     &$3131$   &$2886$  &2861  &2932  &2924  &$-$  \\
  $2P^\prime$  &  $-$               &$3343$   &$-109$     &$3234$   &$2926$  &2877  &3021  &2961 &$-$   \\
  $2^3P_2$     & $-$                &$3399$   &$-168$     &$3230$   &$2971$  &2884  &3012  &2957  &$-$ \\
  $1^3D_1$     & $D_{1}^*(2760)$    &$3131$   &$-371$ &$2759$[$2786$] &$2740$ &2788 &2762  &2817  &$2781\pm22$    \\
  $1D$         & $D_2(2740)$        &$3127$   &$-395$     &$2732$   &$2693$  &2773  &2806  &2816 &$2747\pm6$    \\
  $1D^\prime$   & $-$                &$3192$   &$-389$     &$2803$   &$2789$  &2779 &2850  &2845 &$-$    \\
  $1^3D_3$     & $D_{3}^*(2750)$    &$3160$   &$-404$     &$2756$   &$2719$  &2779  &2863  &2833A  &$2763.1\pm3.2$       \\
  \hline\hline
\end{tabular}
\end{center}
\end{table*}

\begin{table*}
\caption{\label{tab:shift} Mass shift $\Delta M$ (in MeV) of each coupled channel.} 
\begin{tabular}{cccccccccccccc} 
\hline 
\hline 
State       &         &$D\pi$   &$D^*\pi$ &$D\eta$ &$D\eta^\prime$   &$D^*\eta$  &$D^*\eta^\prime$ &$D\rho$ &$D^*\rho$  &$D\omega$ &$D^*\omega$  &Total \\
\hline
$1^1S_0$    &                     &$0$     &$-76$    &$0$    &$-7$    &$0$    &$-12$    &$-55$    &$-144$    &$-18$   &$-48$   &$-360$        \\
$1^3S_1$    &                     &$-29$   &$-63$    &$-2$   &$-6$    &$-4$   &$-9$     &$-43$    &$-197$    &$-14$   &$-66$   &$-434$       \\
$2^1S_0$    &$D_0(2550)^0$        &$0$     &$-48$    &$0$    &$-9$    &$0$    &$-8$     &$-79$    &$-168$    &$-26$   &$-56$   &$-395$    \\
$2^3S_1$    &$D_1^*(2600)^0$      &$-1$    &$-4$     &$-1$   &$-5$    &$-2$   &$-5$     &$-78$    &$-204$    &$-25$   &$-67$   &$-393$      \\    
$1^3P_0$    &$D_0^*(2300)$        &$-14$   &$0$      &$-5$   &$0$     &$-5$   &$0$      &$0$      &$-247$    &$0$     &$-82$   &$-353$     \\ 
$1P$        &$D_1(2420)$          &$0$     &$78$     &$0$    &$-8$    &$0$    &$-11$    &$-71$    &$-172$    &$-24$   &$-57$   &$-421$       \\     
$1P^\prime$  &$D_1(2430)$         &$0$     &$44$     &$0$    &$-7$    &$0$    &$-9$     &$-56$    &$-205$    &$-19$   &$-68$   &$-408$    \\ 
$1^3P_2$    &$D_2^*(2460)$        &$-41$   &$-74$    &$-4$   &$-6$    &$-5$   &$-8$     &$-53$    &$-187$    &$-18$   &$-62$   &$-457$      \\
$2^3P_0$    &$-$                  &$2$     &$0$      &$-1$   &$0$     &$-3$   &$0$      &$0$      &$-213$    &$0$     &$-71$   &$-287$     \\ 
$2P$        &$-$                  &$0$     &$-42$    &$0$    &$-3$    &$0$    &$-7$     &$-35$    &$-26$     &$-12$   &$-8$    &$-132$     \\     
$2P^\prime$  &$-$                 &$0$     &$-25$    &$0$    &$-2$    &$0$    &$-7$     &$-17$    &$-41$     &$-6$    &$-13$   &$-109$    \\  
$2^3P_2$    &$-$                  &$-19$   &$-38$    &$-2$   &$-4$    &$-3$   &$-4$     &$-41$    &$-34$     &$-14$   &$-11$   &$-168$     \\ 
$1^3D_1$    &$D_1^*(2760)^0$      &$12$    &$5$      &$1$    &$-1$    &$-4$   &$-1$     &$-14$    &$-274$    &$-5$    &$-91$   &$-371$     \\
$1D$        &$D_2(2740)^0$        &$0$     &$-46$    &$0$    &$-6$    &$0$    &$-8$     &$-76$    &$-176$    &$-25$   &$-58$   &$-395$     \\ 
$1D^\prime$  &$-$                 &$0$     &$-26$    &$0$    &$-4$    &$0$    &$-7$     &$-60$    &$-204$    &$20$    &$-67$   &$-389$     \\
$1^3D_3$    &$D_3^*(2750)$        &$-25$   &$-50$    &$-3$   &$-5$    &$-3$   &$-6$     &$-56$    &$-179$    &$-18$   &$-59$   &$-404$      \\ 
\hline 
\hline
\end{tabular}
\end{table*}

\begin{table*}
\caption{\label{tab:pro} The two quark and molecule probabilities (in $\%$) of the $D$ meson in the coupled channels framework.} 
\begin{tabular}{ccccccccccccccc} 
\hline 
\hline 
State       &         &$D\pi$   &$D^*\pi$ &$D\eta$ &$D\eta^\prime$   &$D^*\eta$  &$D^*\eta^\prime$ &$D\rho$ &$D^*\rho$  &$D\omega$ &$D^*\omega$  &$P_{molecule}$   &$P_{c\bar{u}}$ \\ 
\hline 
$1^1S_0$    &                     &$0$     &$5.3$    &$0$    &$0.4$    &$0$    &$0.6$    &$3.2$    &$7.7$    &$1.1$   &$2.5$   &$20.8$ &$79.2$       \\
\hline 
\hline
\end{tabular}
\end{table*}

\begin{table*}
\caption{\label{tab:width1} Decay widths of the $2S$ and $1P$-wave states.} 
\begin{tabular}{cccccccccc} 
\hline 
\hline 
Channel            &$2^1S_0$       &$2^3S_1$        &$1^3P_0$      &$1^1P_1$      &$1^3P_1$    &$1^3P_2$  \\ 
                   &$D_0(2550)^0$  &$D_1^*(2600)^0$ &$D_0^*(2300)$ &$D_1(2420)$   &$D_1(2430)$ &$D_2^*(2460)$ \\
\hline 
$D\pi$             &$0$            &$0.1$             &$159$         &$0$           &$0$         &$29$ \\
$D^*\pi$           &$186$          &$90$            &$0$           &$13$          &$162$       &$21$   \\    
$D\eta$            &$0$            &$0$             &$0$           &$0$           &$0$         &$0$   \\ 
$D\eta^\prime$     &$0$            &$0$             &$0$           &$0$           &$0$         &$0$  \\     
$D^*\eta$          &$0$            &$0$             &$0$           &$0$           &$0$         &$0$   \\ 
$D^*\eta^\prime$   &$0$            &$0$             &$0$           &$0$           &$0$         &$0$  \\
$D\rho$            &$0$            &$0$             &$0$           &$0$           &$0$         &$0$  \\ 
$D^*\rho$          &$0$            &$0$             &$0$           &$0$           &$0$         &$0$  \\     
$D\omega$          &$0$            &$0$             &$0$           &$0$           &$0$         &$0$  \\ 
$D^*\omega$        &$0$            &$0$             &$0$           &$0$           &$0$         &$0$  \\ 
$D_0^*(2300)\pi$   &$8$            &$0$             &$0$           &$0$           &$0$         &$0$  \\ 
$D_2^*(2460)\pi$   &$0$            &$0.02$           &$0$           &$0$           &$0$         &$0$  \\ 
$D_1(2420)\pi$     &$0$            &$16$            &$0$           &$0$           &$0$         &$0$  \\ 
$D_1(2430)\pi$     &$0$            &$1$             &$0$           &$0$           &$0$         &$0$  \\
$D_sK$             &$0$            &$1$             &$0$           &$0$           &$0$         &$0$ \\
$D_sK^*$           &$0$            &$0$             &$0$           &$0$           &$0$         &$0$  \\
$D_s^*K$           &$0$            &$4$            &$0$           &$0$           &$0$         &$0$  \\
$D_s^*K^*$         &$0$            &$0$             &$0$           &$0$           &$0$         &$0$  \\
Total              &$194$          &$111$           &$159$         &$13$          &$162$       &$50$  \\
Exp.               &$165\pm24$     &$141\pm23$      &$229\pm16$    &$31.3\pm1.9$  &$314\pm29$  &$47.3\pm0.8$ \\
\hline 
\hline
\end{tabular}
\end{table*}

\begin{table*}
\caption{\label{tab:width2} Decay widths of the $2P$ and $1D$ charmed mesons.} 
\begin{tabular}{ccccccccc} 
\hline 
\hline 
Channel            &$2^3P_0$    &$2P$    &$2P^\prime$    &$2^3P_2$    &$1^3D_1$        &$1D$           &$1D^\prime$ &$1^3D_3$ \\ 
                   &$-$         &$-$     &$-$            &$-$         &$D_1^*(2760)^0$ &$D_2(2740)^0$  &$-$         &$D_3^*(2750)$ \\
\hline 
$D\pi$             &$27$        &$0$     &$0$            &$32$        &$16$            &$0$            &$0$         &$35$     \\
$D^*\pi$           &$0$         &$17$    &$0.2$          &$30$        &$2$            &$58$           &$46$        &$39$    \\    
$D\eta$            &$0$         &$0$     &$0$            &$0$         &$0$             &$0$            &$0$         &$0$    \\ 
$D\eta^\prime$     &$0$         &$0$     &$0$            &$0$         &$0$             &$0$            &$0$         &$0$    \\     
$D^*\eta$          &$0$         &$0$     &$0$            &$0$         &$0$             &$0$            &$0$         &$0$    \\ 
$D^*\eta^\prime$   &$0$         &$0$     &$0$            &$0$         &$0$             &$0$            &$0$         &$0$    \\
$D\rho$            &$0$         &$28$    &$21$           &$15$        &$1$            &$118$          &$30$        &$5$   \\ 
$D^*\rho$          &$65$        &$164$   &$27$           &$88$        &$2$             &$0$            &$17$        &$0$  \\     
$D\omega$          &$0$         &$10$    &$7$            &$5$         &$0.4$            &$38$           &$10$        &$2$   \\ 
$D^*\omega$        &$21$        &$56$    &$10$           &$31$        &$0$             &$0$            &$4$         &$0$    \\
$D_0^*(2300)\pi$   &$0$         &$4$     &$5$            &$0$         &$0$             &$2$            &$4$         &$0$    \\ 
$D_2^*(2460)\pi$   &$0$         &$55$    &$7$            &$36$        &$0.01$             &$1$            &$148$       &$2$   \\
$D_1(2420)\pi$     &$87$        &$2$     &$0$            &$60$        &$123$           &$2$            &$2$         &$1$    \\ 
$D_1(2430)\pi$     &$0.1$       &$4$     &$9$            &$5$         &$4$             &$1$            &$2$         &$1$\\
$D_sK$             &$14$        &$0$     &$0$            &$4$         &$33$            &$0$            &$0$         &$3$  \\
$D_sK^*$           &$0$         &$25$    &$10$           &$5$         &$0$             &$0$            &$0$         &$0$   \\
$D_s^*K$           &$0$         &$2$     &$12$           &$0.1$       &$1$            &$1$            &$42$        &$1$  \\
$D_s^*K^*$         &$0$         &$61$    &$104$          &$88$        &$0$             &$0$            &$0$        &$0$  \\
Total              &$213$       &$428$   &$211$          &$399$       &$182$           &$221$          &$304$       &$89$\\
Exp.               &$-$         &$-$     &$-$            &$-$         &$180\pm40$      &$88\pm19$      &$-$         &$66\pm5$\\ 
\hline 
\hline
\end{tabular}
\end{table*}

\section{Summary}
\label{sec:summary}
In recent years, the gradually enriched experimental results of the charmed mesons have provided an important platform for exploring the coupled channel effects in the mass spectrum of the charmed mesons. In this work we have systematically calculated the mass spectrum of the charmed mesons with the nonrelativistic quark model by involving the coupled channel effects. Our results are in reasonable agreement with empirical results and other quark model predictions. The strong decay widths are further calculated within the $^3P_0$ model. When calculating the mass shifts and widths by the $^3P_0$ model, we adopt the realistic wave functions obtained from the potential model to account for the differences in the space distribution of different meson. We summarize the main points as follows.

\begin{itemize}
    \item 
    By considering the unquenched coupled channel effects, we have given a reasonable description of the masses of the charmed mesons, and analyzed the strong decay widths on the basis of these masses by considering the $S-D$ mixing.
    \item 
    We predict that the $D$ meson have $79.2 \%$ $c\bar{u}$ component and $20.8 \%$ coupled channel components, which include  $D^*\pi$ $5.3 \%$, $D\rho$ $3.2 \%$, $D^*\rho$ $7.7 \%$, $D^*\omega$.  
    \item 
    For the $1P$-wave states,  the $D_0^*(2300)$, $D_1(2420)$ and $D_1(2430)$ are difficult to be assigned  as the $1^3P_0$, $1P$ and $1P^\prime$, respectively, which implies that these state could have more complicate structure. 
    \item 
    The $D_2^*(2460)$ is a good candidate for $1^3P_2$ state. we have effectively explained the masses and decay properties of the $D_1^*(2600)$ and $D_1^*(2760)$ by considering the $S$-$D$ mixing.
 
    \item 
    We also predicted the the masses and widths of the $2P$ states without experimental information such as the $2P$-wave states, which will be helpfully to observed them in future.
\end{itemize}

\section{Acknowledgements}

This work is partly supported by the National Key R\&D Program of China under Grant No. 2024YFE0105200, and by the Natural Science Foundation of Henan under Grant No. 232300421140 and No. 222300420554, the National Natural Science Foundation of China under Grant No. 12475086 and No. 12192263.

\bibliographystyle{unsrt}
\bibliography{cite}  

\end{document}